\documentclass{article}

\usepackage{arxiv}

\usepackage[utf8]{inputenc} % allow utf-8 input
\usepackage[T1]{fontenc}    % use 8-bit T1 fonts
\usepackage{hyperref}       % hyperlinks
\usepackage{url}            % simple URL typesetting
\usepackage{booktabs}       % professional-quality tables
\usepackage{amsfonts}       % blackboard math symbols
\usepackage{nicefrac}       % compact symbols for 1/2, etc.
\usepackage{microtype}      % microtypography
\usepackage{lipsum}		    % Can be removed after putting your text content
\usepackage{graphicx}
\usepackage{multirow}
\usepackage[numbers,square]{natbib}
\usepackage{doi}
\usepackage{amsmath}
\usepackage[bottom]{footmisc}
\usepackage{footnote}
\hypersetup{hidelinks}
\usepackage{lipsum}

\title{PriorPath: Coarse-To-Fine Approach for Controlled De-Novo Pathology Semantic Masks Generation}

\date{}


\author{\href{https://orcid.org/0000-0002-0939-3379}{\includegraphics[scale=0.06]{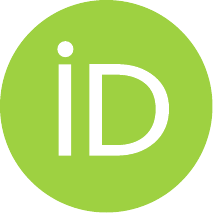}\hspace{1mm}Nati Daniel$^{1}$} \\
    Technion-IIT, Israel\\
		\And
  \href{https://orcid.org/0009-0005-5937-2727}{\includegraphics[scale=0.06]{orcid.pdf}\hspace{1mm}May Nathan$^{2}$} \\
	Technion – IIT, Israel\\
		\And
  \href{https://orcid.org/0009-0002-2300-4365}{\includegraphics[scale=0.06]{orcid.pdf}\hspace{1mm}Eden Azeroual$^{2}$}\\
    Technion – IIT, Israel\\
    \And
  \href{https://orcid.org/0000-0003-3304-4854}{\includegraphics[scale=0.06]{orcid.pdf}\hspace{1mm}Yael Fisher$^{3}$}\\
    Rambam Health Care Campus, Israel\\
	    \And
     \href{https://orcid.org/0000-0002-5345-8491}{\includegraphics[scale=0.06]{orcid.pdf}\hspace{1mm}Yonatan Savir$^{1,2,}$\thanks{Corresponding author, e-mail: \href{mailto:yoni.savir@technion.ac.il}{yoni.savir@technion.ac.il}. $^{1}$Department of Physiology, Biophysics, and Systems Biology, Faculty of Medicine, Technion - IIT, Haifa, Israel. $^{2}$Network Biology Research Lab, Faculty of Electrical and Computer Engineering, Technion – IIT, Haifa, Israel $^{3}$ Division of
Pathology, Rambam Health Care Campus, Haifa, Israel.
     }} \\
	Technion-IIT, Israel
}

\renewcommand{\headeright}{Daniel N et al.}
\renewcommand{\shorttitle}{PriorPath: Controlled De-Novo Pathology Semantic Masks Generation}

\begin{document}

\maketitle

\begin{abstract}
Incorporating artificial intelligence (AI) into digital pathology offers promising prospects for automating and enhancing tasks such as image analysis and diagnostic processes. However, the diversity of tissue samples and the necessity for meticulous image labeling often result in biased datasets, constraining the applicability of algorithms trained on them. To harness synthetic histopathological images to cope with this challenge, it is essential not only to produce photorealistic images but also to be able to exert control over the cellular characteristics they depict. Previous studies used methods to generate, from random noise, semantic masks that captured the spatial distribution of the tissue. These masks were then used as a prior for conditional generative approaches to produce photorealistic histopathological images. However, as with many other generative models, this solution exhibits mode collapse as the model fails to capture the full diversity of the underlying data distribution. In this work, we present a pipeline, coined PriorPath, that generates detailed, realistic, semantic masks derived from coarse-grained images delineating tissue regions. This approach enables control over the spatial arrangement of the generated masks and, consequently, the resulting synthetic images. We demonstrated the efficacy of our method across three cancer types, skin, prostate, and lung, showcasing PriorPath's capability to cover the semantic mask space and to provide better similarity to real masks compared to previous methods. Our approach allows for specifying desired tissue distributions and obtaining both photorealistic masks and images within a single platform, thus providing a state-of-the-art, controllable solution for generating histopathological images to facilitate AI for computational pathology.
\end{abstract}


\section{Introduction}\label{intro}
The increasing use of digital and computational biology, along with the incorporation of AI and machine learning, is significantly impacting the field of medicine \citep{Kiehl2022}. By analyzing vast amounts of biological data, these technologies enhance diagnosis accuracy, optimize treatment assignment, and pave the way for personalized medicine \citep{hart2019classification, tosta2017segmentation, cui2019deep, wang2016deep, kovalev2016deep, fakoor2013using, geras2017high, djuric2017precision, larey2022harnessing, czyzewski2021machine, daniel2022deep, janowczyk2016deep, shen2017deep}. AI algorithms help pathologists process and interpret large volumes of data more quickly, contributing to faster and more precise diagnoses \citep{daniel2022deep}. Moreover, AI can be used to identify patterns and anomalies that lead to inferences of new biomarkers \citep{larey2022fron} and generate predictive insights \citep{czyzewski2021machine}. 

Yet, the development and deployment of AI applications are limited by data availability. Specifically, in computational pathology, the shortage of verified, unbiased, datasets compromises the accuracy and generalizability of AI algorithms in unraveling the complexities of histopathological images and enabling precise diagnostic decision-making \citep{serag2019translational, tizhoosh2018artificial}. This data scarcity serves as the driving force behind the generation of synthetic histopathological images \citep{quiros2019pathologygan, daniel2023between, chen2020generative, guibas2017synthetic, li2022high} by using Generative Adversarial Networks (GANs) \citep{goodfellow2020generative}.

There are two main approaches to generating histopathological images. The first is to generate images from a noise \citep{goodfellow2020generative}. The second uses conditional GANs (cGANs) that receive some prior as in input, particularly cGANs that receive a semantic label mask as an input and then translate it to an image \citep{mirza2014conditional}. The first approach can produce an unlimited number of synthetic images but lacks the ability to control the distribution of cellular features directly. The second approach provides the ability to control the cellular features of the synthetic images precisely, but its scalability is bounded by the quantity and quality of the semantic masks. Manually producing highly detailed synthetic tissue masks is equally impractical. These limitations highlight the importance of producing high-quality synthetic masks, therefore underlining the need within histopathology to address this challenge by generating fine-grained semantic masks that closely replicate the distribution of real data.

\begin{figure*}[t!]
\centering
    \includegraphics[width=\textwidth]{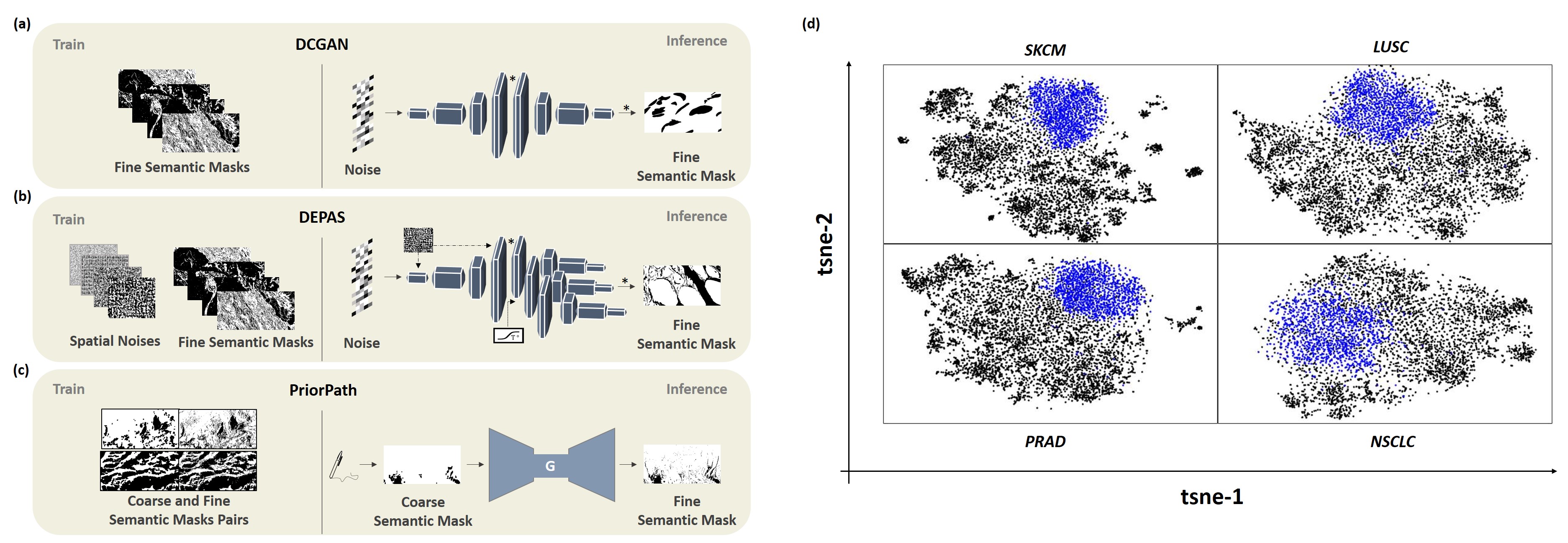}
    \caption{Analysis and Illustration of Generative Approaches for High-Resolution Binary Semantic Masks of Tissue Structure. (a) DCGAN approach: Generates synthetic histopathological fine semantic masks from random noise, requiring real histopathological semantic masks for discriminator processing during training. (b) DEAPS method (current SOTA): Enhances DCGAN with discrete adaptive block final activation, multi-scale discriminators, and spatial noise within hidden layers. (c) PriorPath approach (this work): Produces complex, realistic semantic masks from coarse-grained semantic masks of regions of interest. (d) Demonstration of DEPAS mode collapse: t-SNE projection of Inception’s representations for the DEPAS method (blue) and the Real masks (black). The dataset includes three cancer types with H\&E staining: Skin Cutaneous Melanoma (SKCM), Prostate Adenocarcinoma (PRAD), and Lung Squamous Cell Carcinoma (LUSC), along with a Non-small cell lung carcinoma (NSCLC) dataset with PD-L1 immunohistochemistry.}
    \label{f:f0_1}
\end{figure*}

Various net architectures, such as the Deep Convolutional Generative Adversarial Network (DCGAN) \citep{radford2015unsupervised} and De-novo Pathology Semantic Masks using a generative adversarial model (DEPAS) \citep{larey2023depas}, have been developed for synthetic mask generation by sampling noise from a given distribution. DCGAN is a common architecture model for creating de-novo semantic masks (Fig.~\ref{f:f0_1}a), but the model’s results are repetitive semantic masks and, as such, are not fully for histopathology \citep{larey2023depas}. DEPAS solves this limitation by introducing three main novel additions: injecting spatial noise into each hidden layer in the generator network, adapting multi-scale discriminators, and using a discrete adaptive block to generate synthetic fine tissue masks (Fig.~\ref{f:f0_1}b). Nevertheless, despite the unbiased wide distribution of the training set within the entire mask space, the model exhibits some mode collapse (Fig.~\ref{f:f0_1}d). That is, synthetic masks do not cover the entire physiological semantic mask space. 

Here, to overcome this challenge, we developed a novel approach based on Image-to-Image (I2I) Translation architecture, coined PriorPath, that is able to generate complex, realistic synthetic masks based on a coarse grain spatial distribution of features. We condition the model with manually drawn coarse semantic masks instead of just generating them from noise.  These coarse-grain masks are used to generate a fine-grained mask that can then in turn be used to generate more diverse and realistic synthetic histopathological images. This provides the pathologists with an interactive control of the general visual features of the resulting image. Hence, our approach combines scalability, realistic semantic labeling, and controllability (Fig.~\ref{f:f0_1}c).

\section{MATERIALS AND METHODS} \label{methods}
\subsection{Study population and dataset}\label{dataset_materials}
In this study, we use four histopathology imaging datasets with two different staining techniques. Hematoxylin and eosin (H\&E) histopathological images derived from three specific cancer types: Prostate Adenocarcinoma (PRAD), Skin Cutaneous Melanoma (SKCM), and Lung Squamous Cell Carcinoma (LUSC). These data sets are sourced from the Cancer Genome Atlas (TCGA) research network \citep{tomczak2015review}, where only the imaging information was utilized for all data types. The second staining technique, known as immunohistochemistry (IHC), involved the collection of histopathological images from patients diagnosed with non-small cell lung carcinoma (NSCLC). This  data originated from a study focused on Immune Checkpoint inhibitor therapy, wherein the identification of programmed death-ligand 1 (PD-L1) in tissue biopsies was crucial for treatment determination \citep{wang2016pd}. All Whole Slide Images (WSIs) were obtained from NSCLC patients who underwent a biopsy at the Rambam Health Care Campus. All procedures performed in this study and involving human participants followed the ethical standards of the institutional research committee of the Rambam Medical Center, approval 0522-10-RMB, and with the Declaration of Helsinki of 1964 and its subsequent amendments or comparable ethical standards.

Across all instances, the whole slides, 50 for each H\&E staining, and 55 for IHC PD-L1, were partitioned into patches with dimensions of 512 x 1024 pixels. Patches predominantly comprised of background elements exceeding 85\% coverage were excluded from the analysis. Consequently, a total of $\sim6559-6983$ images were utilized from each H\&E dataset, while 5500 images were extracted from the IHC dataset. In all instances, we generated ground truth for binary tissue semantic masks by converting the patches to grayscale, subsequently applying a high threshold to extract air pixels. The optimal thresholds identified to distinguish between tissue and air pixels were determined to be 204 and 235, within the range of 0-255, for H\&E and IHC, respectively. Table \ref{data_tab} summarizes the main parameters of all the instances, including the splitting information that was employed for training and evaluation purposes.

\begin{table}[!htbp]
    \begin{center}
    \caption{Main parameters of all four datasets used in the training and evaluation processes.}
    \label{data_tab}
        \begin{tabular}{l c c c c c} \\
			\hline
             Parameter / Dataset & PRAD \citep{tomczak2015review} & SKCM \citep{tomczak2015review} & LUSC \citep{tomczak2015review} & NSCLC \citep{daniel2023between}\\
			\hline
			\hline
           	\# of slides & 50 & 50 & 50 & 55\\
                \hline
                \# of participants & 50 & 50 & 50 & 4\\
                \hline
                \# of train patches & 5,983 & 5,759 & 5,559 & 4,000 \\
                \hline
                \# of test patches &  1,000 &  1,000 &  1,000 & 1,500 \\
                \hline
                Patch size &  512 x 1024 &  512 x 1024 &  512 x 1024 & 512 x 1024 \\
                \hline
                Staining method &  H\&E &  H\&E &  H\&E & IHC \\
                \hline
                 Gender &  Male &  Male \& Female &  Male \& Female & Male\\
                \hline
                Year &  2015 &  2015 &  2015 &  2019\\
			\hline
            \hline
		\end{tabular}
	\end{center}
\end{table}

\subsection{Image quality assessment metrics} \label{img_q_metrics}
To quantize the performance of tissue mask generation, we employed three different similarity distance metrics such as Frechet inception distance (FID) \citep{FID_metric}, Kolmogorov–Smirnov (K-S) test \citep{KS_metric}, and Kullback–Leibler (K-L) divergence \citep{KL_metric}.

\subsection{Generating Paired Coarse-Fine Grain Masks for Training} \label{paired_cfg_masks}
To be able to train I2I models \citep{p2p_paper, cyclegan_paper} for various cancer types described in \citep{tomczak2015review}, we had to create paired coarse and fine grains semantic masks. The way we handled it, is to execute over the ground truth of binary tissue semantic masks (explained in Subsection \ref{dataset_materials}). Then, for each fine-grain semantic mask, we applied a set of morphological operations to create its pair of coarse-grain semantic mask. In particular, we performed a two-stage morphological process; the first stage is an erosion followed by a dilation, with an opening kernel in size of 5 x 5 pixels. The second stage is a dilation followed by an erosion, with a closing kernel in size of 10 x 10 pixels.

\subsection{Labeling Manual-made Synthetic Semantic Masks for Evaluation}
One of the main goals of our study was to equip pathologists with a semantic tool enabling the creation of RGB tissue images with high controllability. Hence, we suggest generating images based on manually made coarse masks. Pathologists have extensive background knowledge regarding natural tissue structure. To compensate for our limited familiarity with the basic formation of tissues, as a starting point, the raw coarse rudimentary masks were manually annotated on top of the original RGB tissue images, using an instrument named Labelme \citep{labelme_2023}.

We aimed to create high-quality, semantically richer representations of biological tissue masks. To achieve this, we developed a method that generates synthetic fine-grained semantic masks that closely match the distribution of real masks obtained from RGB tissue images. To learn the real mask distribution, we applied K-means - an iterative algorithm that classifies data into clusters based on centroids and latent distance. For optimal outcomes, we tuned the hyper-parameter that fixates the number of clusters by visual evaluation. Using K-means clustering, we grouped the data based on the prominent visual similarities shared by the masks within each cluster. Then, based on these clusters, we manually drew on an iPad and a smartphone using the Madibeng \citep{madibeng} and Samsung PENUP \citep{samsung_pen} digital platforms, where the art of painting was carried out with varying levels of accuracy; consequently, the results represent trustworthy scribbles of individual pathologists.

Eventually, this resulted in about 100 raw binary sketches per cancer type. To demonstrate the scalability of our approach, we applied extensive augmentations to the masks, generating a large repository of coarse semantic masks for each cancer type. Over these outputs, we performed additional processing, such as binarization and resizing.

\subsection{The CycleGAN and pix2pix formulation} \label{p2p_cg_section}
In this work, we have integrated both pix2pix \citep{p2p_paper} and CycleGAN \citep{cyclegan_paper} in our framework. Both are cGAN frameworks \citep{mirza2014conditional} for I2I translation tasks, to generate synthetic histopathological masks that achieve state-of-the-art (SOTA) results of fine geometry / pattern mask details and realistic textures. The architecture of pix2pix and CycleGAN is based on the GAN paradigm with one main modification of conditional input to the generator's loss function. The pix2pix architecture includes a generator network, which is based on an Encoder-Decoder architecture, a U-Net \citep{frangi2015medical} (like models with skip connections), that aims to learn the deep representation of the input mask and then decode it. In addition, it includes a part of the discriminator network, which is based on PatchGAN that takes a patch mask $N \cdot N$ and predicts for every pixel of the patch whether it belongs to a real mask or a synthetic mask. In particular, the inputs of the pix2pix generator, $G$, architecture are: noise $z$, coarse grain mask of size 512 x 1024 pixels, $x$. $y$ is the synthetic fine-grain mask output of size 512 x 1024 pixels, such that $x$ and $y$ are data pairs. In Discriminator, $D$, architecture, we take the pair of ordered coarse and synthetic fine grain masks, $x$ and $y$, and try to train the network to distinguish between them. As a result, the $G$ is forced to study the true realistic distribution of information and to reduce the loss from the discriminator.
Thus, it obtains higher-quality masks even in the smallest geometry/pattern details.

Hence, the objective of the pix2pix model is expressed as:

\begin{equation}
\min_{G}\max_{D}\mathbb{L}_{cGAN}[(G,D)] + \lambda \cdot \mathbb{L}_{L1}[(G)] \label{eq_p2p_gan_loss}
\end{equation}

where $G$ is a generator, $D$ is a discriminator, and $\lambda$ represents a regularization parameter.

$\mathcal{L}_{cGAN}(G,D)$ is conditional GAN loss (\ref{eq_cgan}), $\mathcal{L}_{L1}(G)$ is a L1 distance loss (\ref{eq_l1}).

\begin{equation}
\mathbb{E}_{x,y}[\log{D(x,y)}] +  \mathbb{E}_{x,z}[\log{(1 - D(x,G(x,z))}] \label{eq_cgan}
\end{equation}

\begin{equation}
\mathbb{E}_{x,y,z}[||y - G(x,z)||_1] \label{eq_l1}
\end{equation} 

where $y$ represents the fine grain mask, $x$ is the coarse grain mask, and $G(x,z)$ is the generated mask given the prior $x$.

In CycleGAN, the used data is unpaired (i.e., without having explicit pairs), unlike pix2pix. Therefore, there may not be a meaningful translator learner that can take a pixel from one mask and convert that into another pixel in a second mask.
But rather it learns a cycle representation instead. CycleGAN architecture includes two generator networks, $G$ and $F$, and two discriminators, $D_{X}$ and $D_{Y}$: where the first generator, $G$, takes an input coarse grain mask $x$ and generates a synthetic fine grain mask $y$, while the second generator, $F$, does the vice versa process.
The $D_{X}$ and $D_{Y}$ verify that the input grain masks generated by the $G$ and $F$, are in the same distribution as the masks of the $y$ and $x$ domains, respectively, consistent with the AKA cycle.

Hence, the objective of the CycleGAN model is expressed as:
\begin{equation}
\min_{G,F}\max_{D_{X},D_{Y}}\mathbb{L}[(G,F,D_{X},D_{Y})] \label{eq_cyclegan_loss}
\end{equation}

where $\mathbb{L}[(G,F,D_{X},D_{Y})]$ (\ref{eq_cyclegan_loss_full}) is a combination of GAN losses (\ref{eq_GDYX}) and Cycle loss (\ref{eq_cyc_loss}).

\begin{equation}
\begin{split}
&\mathbb{L}_{GAN}[(G,D_{Y},X,Y)] + \\
&\mathbb{L}_{GAN}[(F,D_{X},Y,X)] + \\
&\lambda \cdot \mathbb{L}_{cyc}[(G,F)] \label{eq_cyclegan_loss_full}
\end{split}
\end{equation}

where $\lambda$ represents a regularization parameter.

\begin{equation}
\begin{split}
&\mathbb{E}_{y}[\log{D_{Y}(y)}] + \mathbb{E}_{x}[\log{(1 - D_{Y}(G(x))}], \\
&\mathbb{E}_{x}[\log{D_{X}(x)}] + \mathbb{E}_{y}[\log{(1 - D_{X}(G(y))}]
\end{split}
\label{eq_GDYX}
\end{equation}

\begin{equation}
\mathbb{E}_{x}[||F(G(x)) - x||_1] + \mathbb{E}_{y}[||G(F(y)) - y||_1] \label{eq_cyc_loss}
\end{equation}

where $y \in Y$ represents the fine grain mask domain, and $x \in X$ is the coarse grain mask domain.

\subsection{Coarse-to-Fine Mask Generation Approach}
To enhance controllability, the generative process of producing synthetic tissue fine masks is initialized by specifying a prior of desired coarse-grained semantic masks representing tissue and applying it to the model (Image Translation). The mechanism is based on the cGAN architecture and includes both paired and unpaired models (described in subsection \ref{paired_cfg_masks}). We utilized and trained these models on our paired coarse-fine grain masks (subsection \ref{paired_cfg_masks}). On the one hand, explicit supervision \citep{p2p_paper} during learning is less flexible and therefore more limiting, but on the other hand, it can be expected to provide more precise results. Alternatively, the \citep{cyclegan_paper} can result in more scalable but may produce unexpected outcomes.

Comparison of the primary results of both models clearly confirmed that the pix2pix model \citep{p2p_paper} is more effective at capturing the characteristics of cellular tissue and improving the diversity in the spatial distribution of fine-grained tissue masks. Consequently, we have decided to proceed with the pix2pix model only for further experiments and results analysis.

\subsection{Photorealistic RGB Image Generation Model}
Paired image translation encompasses a set of tasks aimed at transforming images from one domain to another, using input-output image training pairs \citep{isola2017image}. One such task involves the insertion of a semantic map into an image and subsequently translating it, leveraging additional information, such as class labels provided alongside the image during the training phase.

In the subsequent stage of our pipeline, we employed pix2pixHD \citep{wang2018high}. This network facilitated the generation of synthetic histopathological images using the provided fine-grained semantic masks (while directly translating coarse semantic masks to histopathological images yields low-fidelity results \citep{larey2023depas}). The generator component of pix2pixHD comprises convolutional residual layers \citep{Resnet50_paper} and operates on a 512 x 1024 pixels semantic mask input to generate corresponding 512 x 1024 pixels high-resolution RGB images. Furthermore, our approach incorporated two multiscale discriminators, both employing a CNN architecture that operates at distinct image scales.

\subsection{Training Procedure}
We employed the PyTorch framework \citep{Pytorch_framework} and trained the model on a single NVIDIA GeForce RTX A6000 GPU with 48GB GPU memory.

To ensure optimal convergence for the model, we carefully tuned the hyperparameters using the Adam Solver \citep{Adam_paper}. We set beta-1 at 0.5 and beta-2 to 0.999, employed a mini-batch of size 1, and initialized the learning rate to 2e-4. During the first 50 epochs, we maintained a constant learning rate and then linearly decayed it to zero over the subsequent 50 epochs. We initialized the weights from a Gaussian distribution with a mean of 0 and a standard deviation of 0.02. We utilized ReLU activation functions for the generator architecture, while the discriminator architectures employed leaky ReLUs with a slope of 0.2. Additionally, we applied reflection padding, as demonstrated in \citep{p2phd_paper}, to enhance the network's performance.

Our optimization loss function comprises two terms. The first term is the mean square error (MSE) between the discriminator's average predictions for synthetic and real masks. The second term is the classic adversarial loss based on binary cross entropy (BCE), supplemented by two feature-based matching losses. These matching losses aim to enforce the synthetic output mask to resemble the specific real mask, preserving the conditional features. All elements of the loss function were equally weighted with a value of 1.

\begin{figure}[thpb]
\centering
\includegraphics[width=\linewidth]{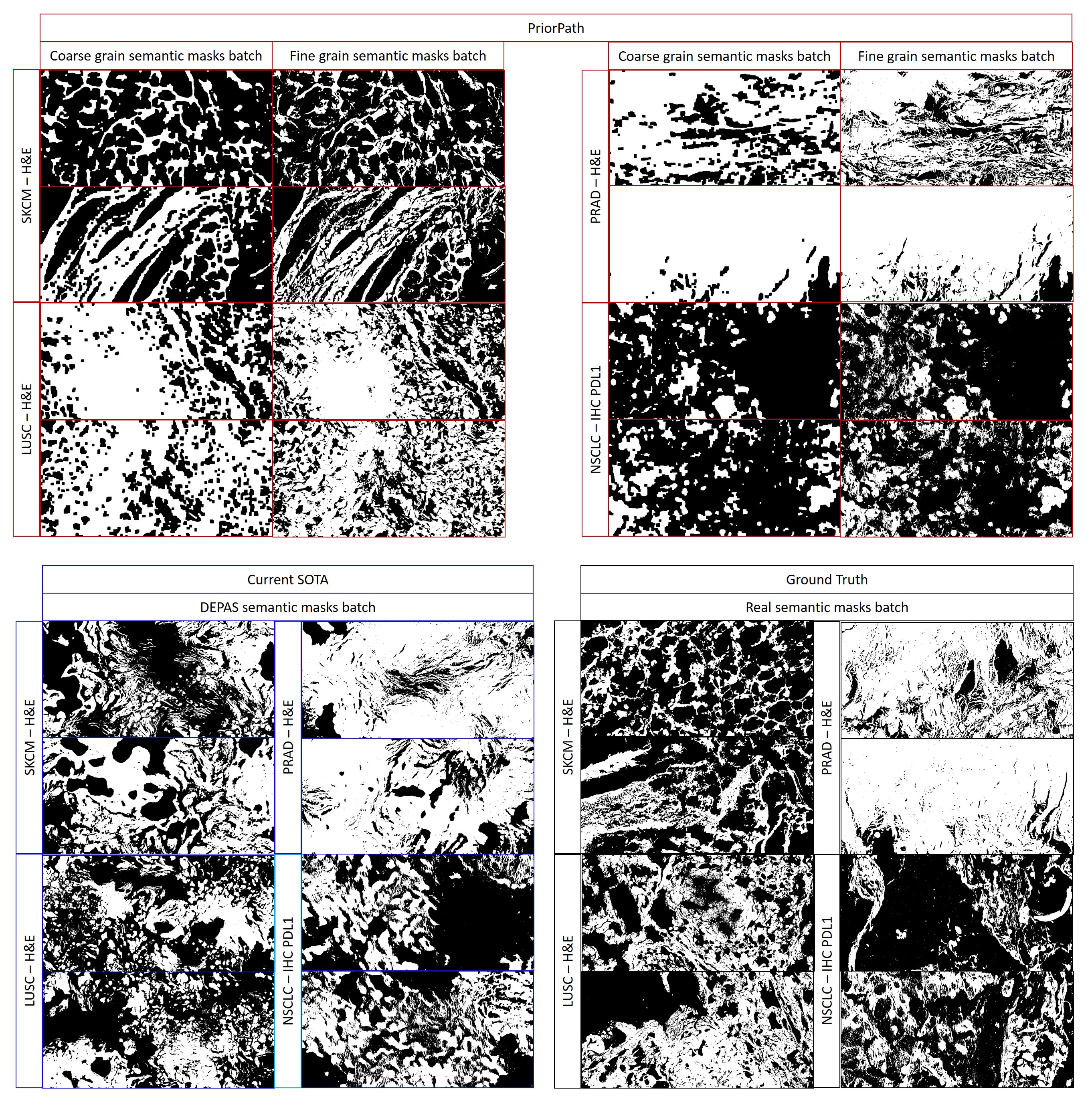}
\caption{Examples of PriorPath tissue mask generation compered with DEPAS. These figures show examples of four types of cancers, and three different organs: skin, prostate, and lung. For each realization, we show PriorPath fine-grain tissue mask generation from coarse-grain masks. The bottom panels show representative tissue masks taken from real biopsy patches, AKA Ground Truth (bottom, right), and by the current SOTA baseline, DEPAS (bottom, left). We show that PriorPath provides tissue masks from a distribution that is closer to the Ground Truth, and is superior in both quality and control of tissue mask generation rather than DEPAS’s outputs (as quantified in Table~\ref{tissue_mask_tab}).}
\label{f:f2}
\end{figure}

\section{RESULTS}\label{results}
\subsection{Coarse-to-fine translation model}
We used two different baseline models for the coarse-to-fine mask approach: CycleGAN \citep{cyclegan_paper} and pix2pix \citep{p2p_paper} (subsection \ref{p2p_cg_section}). Then, we performed quantitative evaluation methodology over the generated fine masks, by calculating the FID score between the synthetic masks stem from pix2pix to the real histopathological masks. For comparison, we performed the same evaluation on the tissue masks generated by CycleGAN for the four different datasets shown in Table~\ref{data_tab}. Specifically, pix2pix's FID scores were better than CycleGAN by a factor of 1.316 for H\&E staining datasets, and by a factor of 1.378 for the IHC staining dataset. Furthermore, qualitative results showed that pix2pix succeeded in learning to generate fine masks compared to CycleGAN, which lacked those fine tissue details and produced relatively coarse output masks. Thus, suggesting that pix2pix gives better control over the spatial distribution of the tissue masks (Fig.~\ref{f:f2}).

\subsection{Synthetic Semantic Tissue Masks}
\begin{figure*}[thpb]
\centering
	\includegraphics[width=\linewidth]{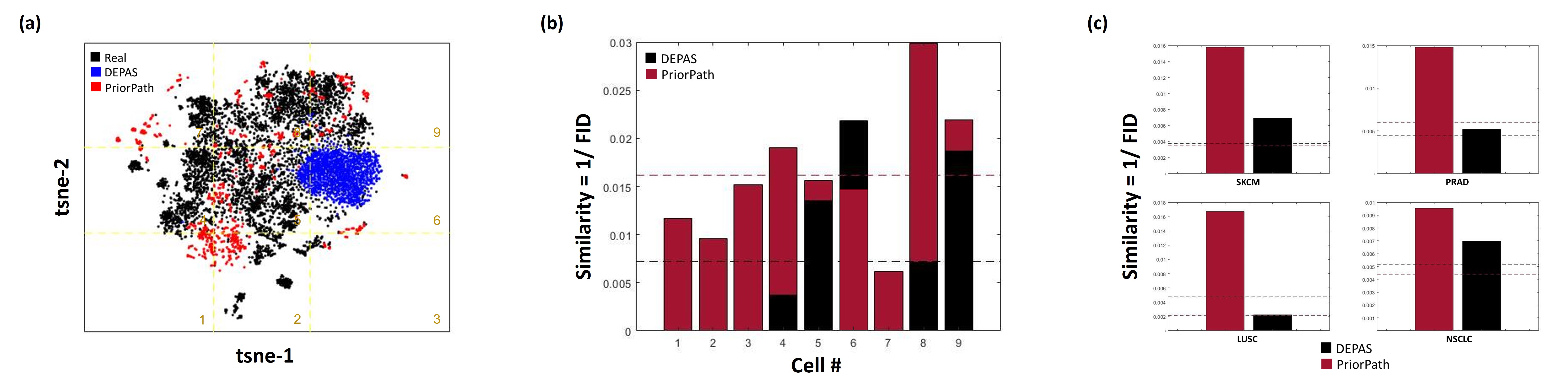}
	\caption{Illustration of PriorPath results, demonstrating the benefits of controlling the distribution while preserving high similarity to the real distribution. (a) A t-SNE projection of inception’s representations for real skin cancer masks (black), DEPAS synthetic masks (blue), and PriorPath synthetic masks (red). PriorPath covers the real pathology mask space. The yellow line defines a grid of size 3 x 3 cells. (b) The similarity metric between the DEPAS and PriorPath synthetic masks and the real ones for each cell inside the grid of 3 x 3. The horizontal lines are the average similarity scores of the synthetic and real masks when taking into account all the masks (in all the grid's cells). (c) The local average similarity over the cell in a grid of size 3 x 3 for all four cancer realizations. The horizontal lines represent the global average similarity scores calculated over all masks (without dividing the representations into the grid).}
	\label{f:f3}
\end{figure*}

We compared PriorPath to the current two SOTA techniques for generating synthetic masks: DCGAN and DEPAS. One of the main limitations of DEPAS is its mode collapse (Fig.~\ref{f:f0_1}d). Fig.~\ref{f:f3}a demonstrates how PriorPath is able to generate masks that cover the real physiological space of real masks. To show that PriorPath masks can cover that semantic phase space without losing their similarity to real ones, we have divided the semantic space into regions and, for each region, calculated the similarity based on FID. Fig.~\ref{f:f3}b shows that for each region, there are PriorPath masks that reside within this region and that their similarity is comparable with that of DEPAS. The overall average local similarity of PriorPath is much better than that of DEPAS. Table~\ref{tissue_mask_tab} summarises the improvement of ProirPath, compared to trained models of DEPAS and DCGAN, in three distances: KL, KS, and FID.

\subsection{Synthetic Photorealistic RGB Images}
\begin{figure}[thpb]
\centering
\includegraphics[width=\linewidth]{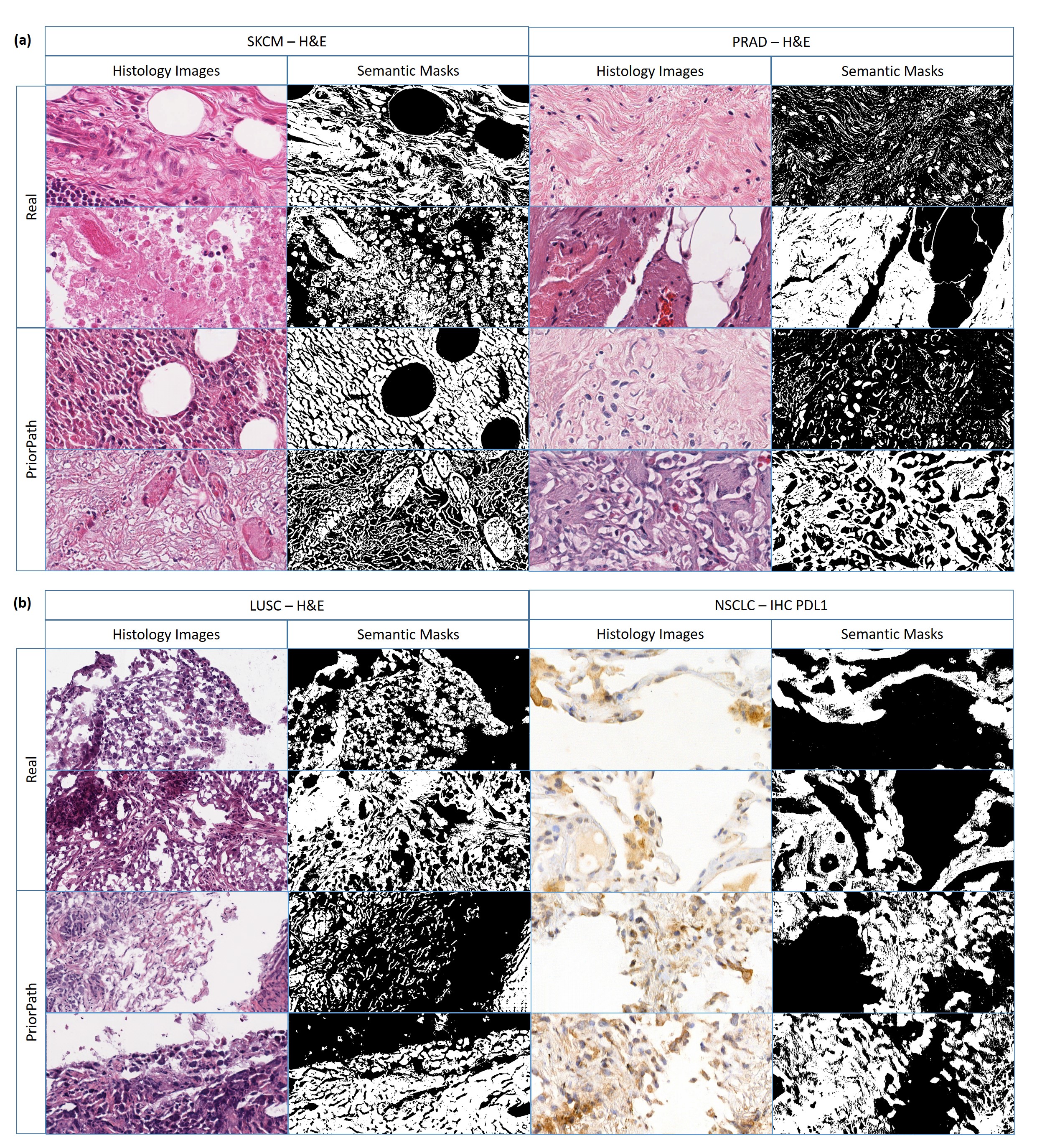}
\caption{Full pipeline photorealistic results. Samples of fine-grain tissue masks and their corresponding histopathological photorealistic synthetic RGB images for all four types of cancer realizations. For each realization, pairs of real and synthetic masks and their corresponding images are shown. (a) Skin prostate tissues. (b) Lung tissues.}
\label{f:f4}
\end{figure}

In addition to the primary image translation stage that utilizes a coarse-to-fine mask approach, we further evaluate the full pipeline in the synthetic photorealistic histopathology perspective, by applying the synthetic tissue fine masks to \citep{wang2018high}, and training over every cancer realization (subsection \ref{dataset_materials}) in a supervised manner. Finally, we compared their outputs to the real histopathological images. Examples of the different tissue fine-grained masks and RGB images for various tissue staining techniques (such as H\&E and IHC) are shown in (Fig.~\ref{f:f4}a-b).

\begin{table}[!htbp]
\begin{center}
\setlength{\tabcolsep}{12pt}
\caption{Similarity metrics between PriorPath de-novo digital pathology semantic masks and current SOTA masks. $^a$, H\&E stained tissues; $^b$, IHC stained tissues. KS, Kolmogorov–Smirnov test; KL, Kullback–Leibler divergence; FID, Fréchet inception distance. $\downarrow$ - symbol indicates lower is better similarity.}
\label{tissue_mask_tab}
        \begin{tabular}{l c c c c c}
			%\toprule
			\hline
		  & & \multicolumn{3}{c}{Image quality metrics}\\
            \cmidrule{3-5}
             Method & Dataset & KS $\downarrow$  & KL  $\downarrow$  & FID $\downarrow$  \\
			%\midrule
			\hline
			\hline
   		   DCGAN \citep{radford2015unsupervised} & & 0.112 & 6.163 & 554.699 \\
     		 DEPAS \citep{larey2023depas} & \multirow{2}{*}{PRAD$^{a}$} & 0.309 & 261.584 & 198.704 \\
        	   PriorPath (ours) & & 0.056 & 0.309 & 71.246 \\
			\hline
            \hline
                DCGAN \citep{radford2015unsupervised} & & 0.251 & 82.569 & 495.516  \\
		      DEPAS \citep{larey2023depas} & \multirow{2}{*}{SKCM$^{a}$} & 0.314 & 234.171 & 138.604\\
        	PriorPath (ours) & & 0.074 & 0.355 & 61.892 \\
        	\hline
            \hline
                DCGAN \citep{radford2015unsupervised} & & 0.214 & 51.578 & 448.568 \\
			DEPAS \citep{larey2023depas} & \multirow{2}{*}{LUSC$^{a}$} & 0.282 & 260.163 & 437.812 \\
        	PriorPath (ours) & & 0.137 & 1.772 & 60.782 \\
        	\hline
            \hline
   		  DCGAN \citep{radford2015unsupervised} & & 0.086 & 0.904 & 424.950 \\
     		DEPAS \citep{larey2023depas} & \multirow{2}{*}{NSCLC$^{b}$} & 0.264 & 185.724 & 138.629 \\
                PriorPath (ours) & & 0.112 & 1.279 & 103.348 \\
            \hline
		\end{tabular}
	\end{center}
\end{table}

To ensure the reliability of the histopathological images generated, we employed a qualitative evaluation methodology at the RGB level. This assessment verified the absence of visual artifacts in the images and confirmed their capture of the histopathological geometry and texture characteristics of the real RGB histopathological images.

\section{DISCUSSION}
One of the primary hurdles in creating synthetic images of tissues is the ability to tightly control the arrangement of features within them. Paired GANs offer a promising avenue for enhancing the quality of synthetic images by incorporating semantic masks that capture the spatial characteristics of the tissue. However, controlling the cellular features in the masks themselves poses a challenge. Unlike other fields such as autonomous vehicles, where there are simulation tools that can provide control over a scene, simulating semantic masks that faithfully represent the intricate biological complexity of the image.

A recent study presented a framework designed to produce high-resolution binary masks representing tissue structure, serving as semantic guidance for image translation models coined DEPAS \citep{larey2023depas}. Although DEPAS synthetic masks are realistic for various pathological realizations and yield photorealistic synthetic images, they exhibit mode collapse and cannot capture the entire space of physiological semantic masks. The reason for that is the masks themself are generated from noise and fall into a particular region of the semantic space. 

In this work, we took a different approach. We aimed to create detailed, realistic semantic masks from a coarse-grain image that outlines the regions in which the tissue should be. This allowed us to control the spatial distribution of the generated masks and, by extension, the generated synthetic images. We showed the ability of this approach to cover the semantic mask space for three types of cancer: skin, prostate, and lung. Not only do the masks generated by PriorPath relieve the mode collapse, but also their similarity to the real masks is better. Our approach can allow pathologists and AI developers to define the desired tissue distributions and get photo-realistic masks and images in one platform. This will facilitate AI development in cases where data is scarce or unbiased. In this work, we focus on generating binary images that capture the distribution of the tissue of the relevant organs (skin, prostate, or lung) as a whole. 

A limitation of the current work is that it generates binary masks. Future work should explore the ability to create multilabel masks that also capture the distribution of single-cell features and not just tissue regions. In addition, future work should explore other conditional generative pipelines such as conditional diffusion models \citep{levine2020synthesis, moghadam2023morphology}.

Overall, this study offers a cutting-edge solution to the demanding task of generating synthetic histopathological images along with their semantic information in a manner that is both scalable and controllable.

\section*{Acknowledgments}
We thank Yael Abuhatsera for her valuable discussions. This work is supported by the Israeli Ministry of Science and Technology (MOST) grant \#2149.

\section*{Data Availability statement}
All data supporting this study's findings are available upon request from the corresponding author.

%\section*{Conflict of Interest}
%The authors declare no conflict of interest related to the work reported in this article.

\bibliographystyle{plainnat}

\end{document}